\documentstyle[preprint,aps]{revtex}
\tightenlines
\def\journal#1, #2, #3#4, #5#6#7#8    {
    {#1~} {#2}  (#5#6#7#8) #3#4}

\def\pra{\journal Phys. Rev. A, }

\def\prl{\journal Phys. Rev. Lett., }

\def\cmp{\journal Comm. Math. Phys., }

\def\npb{\journal Nucl. Phys. B, }
\def\plb{\journal Phys. Lett. B, }

\def\mpl{\journal Mod. Phys. Lett. A, }
\def\ijmp{\journal Int. Jour. Mod. Phys. A, }
\def\jpsj{\journal J. Phys. Soc. Jnp., }
\def\jmp{\journal J. Math. Phys., }

\def\Nb{{I\!\! N}}

\newcommand{\beq}[1]{\begin{equation}\label{#1}}
\newcommand\eeq{\end{equation}}
\newcommand{\ba}[1]{\begin{eqnarray}\label{#1}}
\newcommand{\baa}{\begin{eqnarray}}
\newcommand\ea{\end{eqnarray}}
\newcommand{\bee}{\begin{equation}}
\newcommand{\br}[1]{\overline #1}
\def\nn{\nonumber \\}
\def\l{\lambda}
\def\n{\nu}

\newcommand{\h}{Hamiltonian}

\def\hlf{\frac{1}{2}}

\begin{document}
%\input{hrvzn}
%\draft
%{\Huge{\bf
\title{Dynamical symmetry algebra of the Calogero model}
\author{Larisa Jonke and Stjepan Meljanac
\footnote{e-mail address: 
larisa@thphys.irb.hr \\ 
\hspace*{3cm} meljanac@thphys.irb.hr}} 
\address{Theoretical Physics Division,\\
Rudjer Bo\v skovi\'c Institute, P.O. Box 180,\\
HR-10002 Zagreb, CROATIA}
\maketitle
\begin{abstract}
We study the dynamical symmetry algebra of the $N$-body Calogero model 
describing the structure of degenerate levels and
demonstrate that the algebra is
intrisically polynomial.
We discuss some general properties of an algebra of $S_N$-symmetric
operators acting on the
$S_N$-symmetric subspace of the Fock space for any statistical parameter $\n$.
In the bosonic case ($\n=0$) we find the 
algebra of  generators for every $N$.
For $\n\neq 0$, we explicitly reproduce the finite algebra 
for the 4-particle model,  demonstrating
some general features of our construction.
\end{abstract}
\vspace{1cm}
%\hspace{1.7cm} PACS number(s): 

%\newpage
\section{Introduction}

The Calogero model\cite{c} describes the system of $N$ bosonic particles on a 
line interacting through the inverse square and harmonic potential. 
It is completely 
integrable, in both the classical and quantum case, 
the spectrum is known and the wave-functions are  given implicitly. 
Since the model has connections with a host of physical problems, 
including recent proposal that the superconformal Calogero model provides
 a microscopic description of the 
extremal Reissner-Nordstr\"om  black hole\cite{bh},
there is 
considerable interest in finding the basis set of 
orthonormal eigenfunctions, and the structure of the dynamical algebra 
that characterizes the eigenstates of the system.

A lot of insight has been gained by investigating 
the algebraic properties of the Calogero model in terms of
the $S_N$-extended Heisenberg algebra\cite{all}.
For the periodic version of the model\cite{s}, the orthonormal eigenfunctions 
in terms of Jack polynomials\cite{st} have been constructed\cite{ll}.
In the case on line, two- and three-particles systems have been considered
\cite{2body,3body}. The 
authors of Ref.\cite{2body} have shown that the dynamical symmetry algebra 
of the two-body model is a polynomial generalization of the $SU(2)$ algebra.
The same group of authors has also treated the three-body problem\cite{3body}, 
obtaining the polynomial algebra and the action of its generators on the 
orthonornal basis. 
It has been shown that in the  two-body case the polynomial $SU(2)$ 
algebra can be linearized, but an attempt  to generalize this result to the
$N$-body case has led to $(N-1)$ linear $SU(2)$ subalgebras that operate only on
subsets of the degenerate eigenspace\cite{ind}.

The problem  of constructing of the algebra of symmetric one-particle operators 
for the Calogero model \cite{isakov} resulted in a similar algebraic structure
as that we discuss in Section 3. The algebra constructed in \cite{isakov}
 is infinite, independent of 
particle number and the constant of interaction (statistical parameter), and is
generally not known. 
 
In this letter we consider the problem 
of dynamical algebra of the $N$-body Calogero 
model in a new way. We demonstrate that the algebra in question is 
intrisically polynomial (except in the $N=2$ case). 
We discuss some general properties of the algebra of operators acting on the 
$S_N$-symmetric subspace of the Fock space for any $\n$.
In the bosonic case ($\n =0$) we find the 
algebra of  generators invariant under the $S_N$-permutation group.
For $\n\neq 0$, we discuss in detail the  algebras for $N=3,4$, 
thus explaining 
some general features of our construction.

\section{The Calogero model and the $S_N$-symmetric Fock space}

The Calogero model is defined by the following \h :
\beq 1
H=-\hlf\sum_{i=1}^N\partial_i^2+\hlf\sum_{i=1}^N x_i^2+\frac{\n(\n-1)}{2}
\sum_{i\neq j}^N\frac{1}{(x_i-x_j)^2}.\eeq
For simplicity, we have set $\hbar$, 
the  mass of particles and the frequency of harmonic 
oscillators equal to one. The dimensionless constant $\n$ is 
the coupling constant ($\n >-1/2$) and $N$ is the number of particles.

Let us introduce the following analogs of creation and annihilation 
operators\cite{all}:
\beq 4
a_i^{\dagger}=\frac{1}{\sqrt{2}}(-D_i+x_i),\;
a_i=\frac{1}{\sqrt{2}}(D_i+x_i), \eeq
where 
$$ D_i=\partial_i+\n\sum_{i=j}^N\frac{1}{x_i-x_j}(1-K_{ij})$$
are Dunkl derivatives\cite{D}, and the operator $a_i$ annihilates the vacuum.
The elementary generators $K_{ij}$ of the symmetry group $S_N$ exchange 
labels $i$ and $j$:
\ba 5
&&K_{ij}x_j=x_iK_{ij},\; K_{ij}=K_{ji},\; (K_{ij})^2=1,\nn
&& K_{ij}K_{jl}=K_{jl}K_{il}=K_{il}K_{ij},\;{\rm for}\;i\neq j,\;
i\neq l,\;j\neq l,\ea
and we choose $K_{ij}|0\rangle=|0\rangle$.
One can easily check that the commutators of creation and 
annihilation operators (\ref{4}) are 
\beq 6 
[a_i,a_j]=[a_i^{\dagger},a_j^{\dagger}]=0,\;
[a_i,a_j^{\dagger}]=\left(1+\n\sum_{k=1}^NK_{ik}\right)\delta_{ij}-\n K_{ij}.
\eeq
After performing a similarity transformation on the \h \ (\ref{1}), we 
obtain the reduced \h \
\beq 7
H'=\left(\prod_{i<j}^N|x_i-x_j|^{-\n}\right)H\left(\prod_{i<j}^N|x_i-x_j|^{\n}
\right)
=\hlf\sum_{i=1}^N\{a_i,a_i^{\dagger}\}=
\sum_{i=1}^Na_i^{\dagger}a_i+E_0 ,\eeq
acting on the space of symmetric functions.  The constant $E_0$ is ground-state 
energy  $E_0=N[1+(N-1)\n]/2$.
We restrict the Fock space $\{a_1^{\dagger n_1}\cdots a_N^{\dagger n_N}
|0\rangle\}$ 
to the $S_N$-symmetric subspace 
$F_{\rm symm}$, where ${\cal N}=\sum_{i=1}^Na_i^{\dagger}a_i$ acts as the 
total number operator. Next, we
 introduce the collective $S_N$-symmetric operators
\beq 8
B_n=\sum_{i=1}^Na_i^n, \; n=0,1,\ldots,N,\eeq
where $B_0$ is the constant $N$ multiplied by the identity operator, 
and $B_1$ represents 
the center-of-mass operator (up to the constant $N$).
The complete $F_{\rm symm}$ can be described as $\{B_1^{\dagger n_1}
B_2^{\dagger n_2}\cdots B_N^{\dagger n_N}|0\rangle\}$. 
 Note that $[B_1, B_k^{\dagger}]\neq 0$, hence we wish to construct the 
operators $X_k^{\dagger}$ such that $B_1$ commute with $X_k^{\dagger}$ 
for every $k$ greater 
than one. The general solution of this equation is described by any 
symmetric monomial polynomial $m_{\l}(\bar a_1,\ldots,\bar a_N)=
\sum \bar a_{1}^{\l_1}\bar a_{2}^{\l_2}\cdots \bar a_{N}^{\l_N},$
 where ${\br a}_i=a_i-
\frac{B_1}{N},$
and the sum goes over all distinct permutations of $\{\l_1,\l_2,\ldots,\l_N\}$.
The set $\{\l_1,\l_2,\ldots,\l_N\}$ denotes any partition of ${\cal N}$ 
such that
$\sum_{i=1}^N\l_i={\cal N}$ and $\l_1\geq\l_2\geq\cdots\geq\l_N\geq 0$.
The simplest choice of the $(N-1)$ operators commuting with $B_1$ is
\beq 9 
A_n\equiv \br B_n=\sum_{i=1}^N\left(a_i-\frac{B_1}{N}\right)^n, \;
n=2,\ldots,N .\eeq
For later convenience, it is useful to define $A_0=N\cdot{\bf 1}
\hspace{-0.12cm}{\rm I}$ and $A_1=0$.
The $F_{\rm symm}$ symmetric Fock space is now  $\{B_1^{\dagger n_1}
A_2^{\dagger n_2}\cdots A_N^{\dagger n_N}|0\rangle\}$ and after removing 
$B_1$ it reduces to the Fock subspace  
$\{A_2^{\dagger n_2}
\cdots A_N^{\dagger n_N}\}|0\rangle$. We have reduced the problem to the 
$(N-1)$ Jacobi-type operators\footnote{Note that the algebraic sum of all 
coefficients of homogeneous monomials in any $A_k$ is zero.}.
However, two different 
 states $A_2^{\dagger n_2}\cdots A_N^{\dagger n_N}|0\rangle$ 
and $A_2^{\dagger n_2'}\cdots A_N^{\dagger n_N'}|0\rangle$ are generally 
not orthogonal. For example, $\langle 0|A_3^2A_2^{\dagger 3}|0\rangle=\langle 0|
A_2^3A_3^{\dagger 2}|0\rangle\neq 0$.
The total number operator on $F_{\rm symm}$ splits into 
\ba a
&&{\cal N}={\cal N}_1+{\cal\br N},\; {\cal N}^{\dagger}={\cal N}=\sum_{i=1}^N
a_i^{\dagger}a_i,\nn
&&{\cal N}_1^{\dagger}={\cal N}_1=\frac{1}{N}B_1^{\dagger}B_1, \nn
&&{\cal \br N}^{\dagger}={\cal \br N}={\cal N}-{\cal N}_1
\equiv\sum_{k=2}^Nk{\cal N}_k .\ea
Note that ${\cal N}_k$ are the number operators of $A_k^{\dagger}$ but not of 
$A_k$. Namely, $[{\cal N}_k,A_l^{\dagger}]=\delta_{kl}A_k^{\dagger}$, and 
${\cal N}_k(\cdots A_k^{\dagger n_k}\cdots|0\rangle) =n_k 
(\cdots A_k^{\dagger n_k}\cdots|0\rangle)$ for every $k$ greater than one, but 
${\cal N}_k^{\dagger}\neq {\cal N}_k$. If ${\cal N}_k$ were hermitian, then
the eigenstates $A_2^{\dagger n_2}\cdots A_N^{\dagger n_N}|0\rangle$ would be
orthogonal, and vice versa.

\section{The polynomial algebra ${\cal B}_N(\n)$}

Here we construct the  finite (for finite $N$)
 and closed algebra ${\cal B}_N(\n)$, which 
appears naturally when we calculate the commutators between 
the operators $A_i$ and $A_j^{\dagger}$, defined in Eq.(\ref{9}). 
Let us define $S_N$-symmetric operators
$\br B_{n,m}$:
\beq g 
\br B_{n,m}=\sum_{i=1}^N\bar a_i^{\dagger n}\bar a_i^m=\br B^{\dagger}_{m,n},\;
n,m\in \Nb_0.\eeq
There are $(N+4)(N-1)/2$ algebraically independent operators 
contained in the algebra ${\cal B}_N(\n)$, namely 
$2(N-1)$ operators 
$\br B_{n,0}=A_n^{\dagger},\;{\rm for}\;n=2,3,\ldots,N$ 
and their hermitian conjugates, and 
$N(N-1)/2$ operators $\br B_{n,m},\;{\rm for}\;
n,m\geq 1, n+m\leq N$.  $[N/2]$ of the latter are hermitian operators 
$\br B_{n,n}$.
Note that $\br B_{0,0}=N\cdot{\bf 1}\hspace{-0.12cm}{\rm I}$ and 
$\br B_{1,0}=\br B_{0,1}=0$. One can express the operators  $\br B_{n,m}$ 
for $n+m>N$ in terms of algebraically independent operators $\br B_{n,m}$
with $n+m\leq N$. 
The operators $\br B_{n,m}$ can be represented in the symmetric Fock space:
$$\br B_{n,m}A_2^{\dagger n_2}\cdots A_N^{\dagger n_N}|0\rangle\equiv
\br B_{n,m}(\prod A^{\dagger})^{\cal N}|0\rangle=\sum (\prod A^{\dagger})^
{{\cal N}+n-m}|0\rangle.$$
The symbolical expression $(\prod{\cal O})^k$ denotes a product of
operators ${\cal O}_i$  of the total order $k$ in $a_i(a_i^{\dagger})$.
Generally,
\ba h
&&\br B_{1,1}={\cal\br N},\;
[B_1, \br B_{n,m}]=0,\;[\br B_{1,1}, \br B_{n,m}]=(n-m)\br B_{n,m},\nn
&&[\bar a_i,\br B_{n,m}]=n\left[\bar a_i^{\dagger (n-1)}\bar a_i^m-\frac{1}{N}
\br B_{n-1,m}\right].\ea
One can define operators $L_m=-B_{m+1,1}$ and $L_{-m}=-B_{1,m+1}$, for $m\geq 0$
 satisfying 
the centerless Virasoro  algebra\cite{viras}. 
It is easy to check that the "bar" operators
satify the following commutation relation, for any $\n$:
$$[\br L_m,\br L_n]=(m-n)\br 
L_{m+n}+\frac{1}{N}[(n+1)A_{m+1}^{\dagger}\br L_{n-1}-
(m+1)A_{n+1}^{\dagger}\br L_{m-1}].$$  
In the limit of large $N$, it becomes the centerless Virasoro  algebra.

For simplicity, let us first consider $N$ free harmonic oscillators with $\n=0$.
We start from a commutator of bosonic operators, with the 
center-of-mass coordinate removed, i. e.
$[\bar b_i,\bar b_j^{\dagger}]=\delta_{ij}-1/N.$
Using the following relations:
\ba i
[\bar b_i^m,\bar b_j^{\dagger n}]&=&\sum_{k=1}^{{\rm min}(n,m)}\beta_k(m,n)
\bar b_j^{\dagger (n-k)}\bar b_i^{(m-k)}
\left(\delta_{ij}-\frac{1}{N}\right)^k\nn
&=&\sum_{k=1}^{{\rm min}(n,m)}\beta_k(m,n)(-)^{(k-1)}\bar b_i^{(m-k)}
\bar b_j^{\dagger (n-k)}\left(\delta_{ij}-\frac{1}{N}\right)^k, \ea
where 
$$\beta_k(m,n)=\frac{m!n!}{k!(m-k)!(n-k)!},$$
we find the general ${\cal B}_N(0)$-algebra relation
\ba j
&&[\br B_{m',m},\br B_{n,n'}]=\sum_{k=1}^{{\rm min}(n,m)}\beta_k(m,n)
\left(\frac{1}{N}\right)^k
\left\{\left[(N-1)^k+1\right]
\br B_{m'+n-k,m+n'-k}\right. \nn 
&&-\left. \br B_{m',m-k}\br B_{n-k,n'}+
\sum_{s=1}^{{\rm min}(n-k,m-k)}\beta_s(m-k,n-k)\br B_{m'+n-k-s,m+n'-k-s}
\left(1-\frac{1}{N}\right)^s \right\}\nn
&&-\{m'\leftrightarrow n,n'\leftrightarrow m\}.
\ea
For finite $N$, the r.h.s. of Eq.(\ref{j}) can be written in terms of the 
operators $\br B_{n,m}$ with $n+m\leq N$.
Specially, we find
\ba k
&&[A_m,A_n^{\dagger}]= \sum_{k=1}^{{\rm min}(n,m)}\beta_k(m,n)
\left\{\br B_{n-k,m-k}\left(1-\frac{1}{N}\right)^k+ 
\left(-\frac{1}{N}\right)^k\left[A_{n-k}^{\dagger}A_{m-k}-\br B_{n-k,m-k}
\right]\right\},\nn
&&\left[\br B_{m,m},\br B_{n,n}\right]=-\sum_{k=1}^{{\rm min}(n,m)}\beta_k(m,n)
\left(\frac{1}{N}\right)^k\left[\br B_{m,m-k}\br B_{n-k,n}+
\br B_{n,n-k}\br B_{m-k,m}\right].\ea 
One can obtain another form of the first relation in (\ref{k}) by simply putting
$m'=n'=0$ in Eq(\ref{j}).
Note that 
$$[B_{m',m},B_{n,n'}]=\sum_{k=1}^{{\rm min}(n,m)}\beta_k(m,n)
B_{m'+n-k,m+n'-k}-\{m'\leftrightarrow n,n'\leftrightarrow m\},$$
i. e., it is  a linear $W_{1+\infty}$  algebra\cite{vw,vwN}. 
In the limit $N\rightarrow\infty$
the algebra ${\cal B}_N(0)$, Eq.(\ref{j})
 becomes also the $W_{1+\infty}$  algebra.

For $\n\neq 0$, the structure of the algebra ${\cal B}_N(\n)$ becomes more 
complicated. 
New polynomial terms of the form $(\prod_{\alpha}\br B_{n_{\alpha},m_{\alpha}})$
with $\sum_{\alpha}n_{\alpha}\leq n+m'-1,\;\sum_{\alpha}m_{\alpha}
\leq n'+m-1$ appear on the r.h.s. of the commutation relation (\ref{j}). 
The corresponding coefficients are polynomial in $\n$, 
vanishing when $\n$ goes to zero. The coefficients of the leading terms 
($k=1$ in Eq.(\ref{j}))
do not depend on $\n$, i. e., they are the same  for any $\n $.
For example, for arbitrary $N$ and $\n$ we find:
\ba l
&&\left[A_2,A_n^{\dagger}\right]=2n\br B_{n-1,1}
+n\left(\frac{N-1}{N}\right)A_{n-2}^{\dagger}(n-1+\n N)
+n\n\sum_{i=1}^{n-2}\left(A_{n-2-i}^{\dagger}A_i^{\dagger}
-A_{n-2}^{\dagger}\right),\nn
&&\left[A_3,A_n^{\dagger}\right]=3n\left(\br B_{n-1,2}-
\frac{1}{N}A_{n-1}^{\dagger}A_2\right)+
n\br B_{n-2,1}\left[3(n-1)\left(\frac{N-2}{N}\right)+\n(N+2)\right]\nn &&+
n(N-1)(N-2)A_{n-3}^{\dagger}\left(
\frac{(n-1)(n-2)}{N^2}+\frac{2n-3}{N}\n
+\n^2\right)\nn
&&-n\n\sum_{i=2}^{n-2}A_{n-3}^{\dagger}\left[(n-1)\left(\frac{N-2}{N}\right)+
(n-2)\left(\frac{N-1}{N}\right)-1\right]\nn
&&+ n\n\sum_{i=2}^{n-3}
\left[(n-1)\left(\frac{N-2}{N}\right)+\left(\frac{N-1}{N}\right)(n-2-i)+
\right]
A_{n-3-i}^{\dagger}
A_i^{\dagger}\nn &&+3n\n\sum_{i=2}^{n-2}\left(A_i^{\dagger}\br B_{n-i-2,1}-
\br B_{n-2,1}\right)-n\n^2\sum_{i=0}^{n-2}\sum_{j=0}^{n-3}\left(
A_{n-3-j}^{\dagger}A_j^{\dagger}-A_{n-3}^{\dagger}\right)\nn
&&+n\n^2\left[
\sum_{i=0}^{n-2}\sum_{j=1}^{n-3-i}\left(A_{n-3-i-j}^{\dagger}A_i^{\dagger}
A_j^{\dagger}-A_{n-3-i}^{\dagger}A_i^{\dagger}\right)
-\sum_{i=0}^{n-2}\sum_{j=1}^{n-3}\left(A_{n-3-j}^{\dagger}
A_{j}^{\dagger}-A_{n-3}^{\dagger}\right)\right].\ea
For $N\leq 5$, the above commutation relations have the same structure as for 
bosons with $\n =0$, but with the coefficients depending polynomially on 
$\n$.
For $N=4$, we give the complete ${\cal B}_4(\n)$ algebra. The minimal set of 
operators is $\{A_2,A_3,A_4,\br B_{11},\br B_{12},\br B_{13},\br B_{22}\}$ plus 
hermitian conjugates. 
\baa\label{example}
&&[A_3,A_3^{\dagger}]=
9(\br B_{2,2}-1/4A_2^{\dagger}A_2+(1+2\n)\br B_{1,1}+(1+2\n)(1+4\n)),
\nn &&[A_4,A_4^{\dagger}]=16(\br B_{3,3}-1/4A_3^{\dagger}A_3)+
\br B_{2,2}(36-8\n)+
A_2^{\dagger}A_2(9/2+16\n)+8\n \br B_{1,1}^2,\nn 
&&\hspace{1.7cm}+(56\n^2+64\n+42)\br B_{1,1}+6(1+4\n)(4\n^2+13\n+21/4),\nn
&&[A_3,A_4^{\dagger}]=12(\br B_{3,2}-1/4A_3^{\dagger}A_2)+(18+12\n)\br B_{2,1}
,\nn
&&[A_i,\br B_{1,j}]=i(A_{i+j-1}-1/4A_{j}A_{i-1}),\; i=2,3,4,\; j=1,2,3, \nn
&&[A_2,\br B_{2,i}]=4\br B_{1,i+1}+(3/2+6\n)A_{i},\;i=0,1,2,\nn
&&[A_2,\br B_{3,1}]=6\br B_{2,2}+(9/2+6\n)\br B_{1,1},\nn
&&[A_3,\br B_{2,i}]=6(B_{1,i+2}-1/4 \br B_{1,i}A_2)+(3+6\n)A_{i+1},\;i=0,1,2,\nn
&&[A_3,\br B_{3,1}]=9(\br B_{2,3}-1/4\br B_{2,1}A_2)+(9+6\n)\br B_{1,2},\nn
&&[A_4,\br B_{2,i}]
=8(\br B_{1,i+3}-1/4\br B_{1,i}A_3)+(6+4\n)A_{i+2}+(3/4+4\n)A_2A_i,\;
i=0,1,2,\nn
&&[A_4,\br B_{3,1}]=12(\br B_{2,4}-1/4\br B_{2,1}A_3)+(18-24\n)\br B_{1,3}
+(9/4+12\n)\br B_{1,1}A_2\nn
&&\hspace{1.7cm}+(21/2+26\n+8\n^2)A_2,\nn
&&[\br B_{1,2},\br B_{2,1}]=3\br B_{2,2}+1/4A_2^{\dagger}A_2
+(2-6\n)\br B_{1,1}-\br B_{1,1}^2,
\nn &&[\br B_{1,2},\br B_{2,2}]=2\br B_{2,3}-\br B_{1,1}\br B_{1,2}
+1/2\br B_{2,1}A_2
+(2-6\n)\br B_{1,2},\nn
&&[\br B_{1,2},\br B_{3,1}]=5\br B_{3,2}+1/4A_3^{\dagger}A_2-3/2\br B_{1,1}
\br B_{2,1}+(6-14\n)\br B_{2,1},\nn
&&[\br B_{1,2},\br B_{1,3}]=-\br B_{1,4}-1/2\br B_{1,1}A_3+3/4\br B_{1,2}A_2,\nn
&&[\br B_{1,3},\br B_{2,2}]
=4\br B_{2,4}-3/2\br B_{1,2}^2+1/2\br B_{2,1}A_3-(3/8-4\n)\br B_{1,1}A_2+(6
-14\n)\br B_{1,3},\nn
&&[\br B_{1,3},\br B_{3,1}]=8\br B_{3,3}+1/4A_3^{\dagger}A_3
+(18-24\n)\br B_{2,2}+
8\n A_2^{\dagger}A_2\nn &&\hspace{1.7cm}-9/4\br B_{1,2}\br B_{2,1}+
(-9/8+4\n)\br B_{1,1}^2+(6-21\n+14\n^2)\br B_{1,1}.
\ea
As we have already mentioned, all operators $\br B_{i,j}$ with $i+j>4$ are 
algebraically 
dependent and can be expressed in terms of the minimal set of operators.
For example, $A_5=5/6A_2A_3$, $\br B_{4,1}=1/3A_3^{\dagger}\br B_{1,1}+1/2
A_2^{\dagger}\br B_{2,1}$, for $N\leq 4$.

For $N=3$,  the ${\cal B}_3(\n)$ algebra is in 
full agreement with Ref.\cite{3body}. The exact 
correspondence between $Y_s$ and $J$ 
defined in Ref.\cite{3body} and our operators 
$\br B_{i,j}$ is 
\bee\label{corr}
Y_1=-2A_2^{\dagger},\;Y_{3/2}=\frac{12}{\sqrt{6}}A_3^{\dagger},\;
Y_{1/2}=-2\sqrt{6}\,\br B_{2,1},\;J_3={\cal N}_1,\;J=\hlf{\cal \br N}+
\frac{3}{2}\n.\eeq
The commutations relations which define the ${\cal B}_4(\n)$ algebra are new 
and shed more light on the structure of the ${\cal B}_N(\n)$ algebra 
and its representations. 

Alternatively, the ${\cal B}_N(\n)$ algebra can be constructed by 
grouping the generators into  $sl(2)$-spin multiplets. Note that 
$J_+=1/2A_2^{\dagger},\; J_-=1/2A_2,\; J_0=1/8[A_2,A_2^{\dagger}]$ generate 
the $sl(2)$-algebra. The complete set of generators 
spanning the ${\cal B}_N(\n)$ algebra is given by $(N-1)$  nondegenerate 
spin multiplets with $s=1,3/2,2,\ldots ,N/2$. The unique generator with spin 
 $s$  and projection $s_z$ is defined as $J_{s,s_z}=\sqrt{\frac{(s+s_z)!}
{(s-s_z)!}}
 [\underbrace{J_-,
\cdots,[J_-}_{(s-s_z)},A_{2s}^{\dagger}]\cdots]$,
for all statistical 
parameters $\n$.  Detail of this construction will be presented 
elsewhere\cite{inprep}.

\section{Dynamical symmetry of the Calogero model}

The dynamical symmetry ${\cal C}_N(\n)$ of the Calogero model is defined as 
maximal algebra commuting with the \h \ (\ref{7}). The generators of the algebra
${\cal C}_N(\n)$ act among the degenerate states with fixed energy $E={\cal N}
+E_0,$
${\cal N}$ a non-negative integer. 
Starting from any of degenerate states with energy $E$, 
all other states can be reached by applying generators $X_{i,j}$ of the 
algebra.
Degeneracy appears for ${\cal N}\geq 2$. The vacuum $|0\rangle$ and the 
first excited state $B_1^{\dagger}|0\rangle$ are nondegenerate. 
For ${\cal N}=2$ 
the degenerate states are $B_1^{\dagger 2}|0\rangle$ and $A_2^{\dagger}
|0\rangle$; for ${\cal N}=3$ the degenerate states are 
$B_1^{\dagger 3}|0\rangle$, $B_1^{\dagger}A_2^{\dagger}|0\rangle$ and 
$A_3^{\dagger}|0\rangle$, etc. The number of degenerate states ${\cal N}$ 
is given by partitions ${\cal N}_1,\ldots,{\cal N}_k$ of ${\cal N}$ such that 
${\cal N}=\sum_k k{\cal N}_k$.

Let us choose $(N+4)(N-1)/2$ algebraically independent
 generators $X_{i,j},\;(i+j\leq N)$ in the following way:
\beq m
X_{i,j}=\br B_{i,j}\left(\frac{B_1}{\sqrt{N}}\right)^{(i-j)},\;
X_{j,i}=X_{i,j}^{\dagger}=\br B_{j,i}\left(\frac{B_1^{\dagger}}
{\sqrt{N}}\right)^{(i-j)}, \; i\geq j.\eeq
For example,
\baa
&&X_{i,0}=A_i^{\dagger}\left(\frac{B_1}{\sqrt{N}}\right)^{i},\;
X_{0,i}=A_i\left(\frac{B_1^{\dagger}}{\sqrt{N}}\right)^{i},\nn
&&[X_{i,0},X_{j,0}]=[X_{0,i},X_{0,j}]=0,\; X_{i,i}^{\dagger}=X_{i,i}
=\br B_{i,i}.\ea
The generators $X_{i,i}$ are hermitian but they do not commute, since 
$\br B_{i,i}$'s do not commute (see Eq.(\ref{k})). 
On the other hand, the number 
operators ${\cal N}_k$ (\ref{a}) commute but are not hermitian since the states 
$A_2^{\dagger n_2}\cdots A_N^{\dagger n_N}|0\rangle$ are not mutually 
orhtogonal.
Generally,
\ba r
&&\left[{\cal N}_1,X_{i,j}\right]=-(i-j)X_{i,j},\; \left[{\cal \br N},X_{i,j}
\right]=(i-j)X_{i,j},\nn
&& \left[H,X_{i,j}\right]=\left[{\cal N},X_{i,j}\right]=0,
\; {\rm for\; all\; i,j.}\ea

The general structure of the commutation relations for $i\geq j,\;k\geq l$ is
\beq n
\left[X_{i,j},X_{k,l}\right]=\left[\br B_{i,j},\br B_{k,l}\right]
\left(\frac{B_1}{\sqrt{N}}\right)
^{(i-j)+(k-l)}=\sum\left[\prod_{\alpha}X_{n_{\alpha},m_{\alpha}}
\right] ,\eeq
and for $i>j,\;k<l$
\beq o
\left[X_{i,j},X_{k,l}\right]=\sum\left[\prod_{\alpha}X_{n_{\alpha},m_{\alpha}}
g_{n_{\alpha},m_{\alpha}}({\cal N}_1)\right]
+X_{i,j}X_{k,l}f_{ijkl}({\cal N}_1),\eeq
with the restriction $0\leq \sum m_{\alpha}\leq j+l-1
,\;0\leq \sum n_{\alpha}\leq i+k-1$, and similarly for hermitian conjugate 
relations. The functions $f$ and $g$ 
are generally rational functions of ${\cal N}_1$, with the finite action on all
states.
One can show that for $i\geq j$,
\baa \left(\frac{B_1}{\sqrt{N}}\right)^{i}\left(\frac{B_1^{\dagger}}
{\sqrt{N}}\right)^{j}&=&\left(\frac{B_1}{\sqrt{N}}\right)^{(i-j)}
({\cal N}_1+1)\cdots({\cal N}_1+j)\nn &=&({\cal N}_1+1+i-j)\cdots({\cal N}_1+i)
\left(\frac{B_1}
{\sqrt{N}}\right)^{(i-j)},\nonumber\ea
and
\baa \left(\frac{B_1^{\dagger}}{\sqrt{N}}\right)^{j}\left(\frac{B_1}
{\sqrt{N}}\right)^{i}&=&{\cal N}_1
({\cal N}_1-1)\cdots({\cal N}_1-j+1)\left(\frac{B_1}
{\sqrt{N}}\right)^{(i-j)}\nn &=&\left(\frac{B_1}{\sqrt{N}}\right)^{(i-j)}
({\cal N}_1-i+j)\cdots({\cal N}_1-i+1), \nonumber\ea
and similarly for $i<j$. Now it is easy to see that
\bee \left[\left(\frac{B_1}{\sqrt{N}}\right)^{i},\left(\frac{B_1^{\dagger}}
{\sqrt{N}}\right)^{j}\right]=\sum_{k=1}^{{\rm min}(i,j)}\beta_k(i,j)
\left(\frac{B_1^{\dagger}}{\sqrt{N}}\right)^{(j-k)}\left(\frac{B_1}{\sqrt{N}}
\right)^{(i-k)}.\eeq

For general $N$, we present several typical commutators which demonstrate the 
general structure given by Eqs.(\ref{n},\ref{o}): 
\baa
\left[X_{2,0},X_{0,2}\right]&=& -2\left[2X_{1,1}+(N-1)(1+N\n)\right]
({\cal N}_1+1)({\cal N}_1+2)\nn &+&
2X_{2,0}X_{0,2}\frac{2{\cal N}_1+1}{({\cal N}_1+1)({\cal N}_1+2)},\nn
\left[X_{2,0},X_{0,3}\right]&=&-6X_{1,2}({\cal N}_1+2)({\cal N}_1+3)+
6X_{2,0}X_{0,3}\frac{{\cal N}_1+1}{{\cal N}_1({\cal N}_1-1)},\nn
\left[X_{2,0},X_{2,1}\right]&=&-2X_{3,0},\nn
\left[X_{2,0},X_{1,2}\right]&=&-4X_{2,1}({\cal N}_1+1)+2X_{2,0}X_{1,2}
\frac{1}{{\cal N}_1+1},\nn
\left[X_{4,0},X_{3,1}\right]&=&-4(X_{6,0}-\frac{1}{N}X_{3,0}^2),\nn
\left[X_{4,0},X_{0,2}\right]&=&-8X_{3,1}-4\left[3\left(\frac{N-1}{N}\right)+
\n(2N-3)\right]X_{2,0}({\cal N}_1+1)({\cal N}_1+2)\nn &+&3X_{4,0}X_{0,2}
\frac{2{\cal N}_1+1}{({\cal N}_1+1)({\cal N}_1+2)}.\ea

The ${\cal C}_N(\n)$ algebra is intrinsically polynomial.  
For $N=2$, the ${\cal C}_2(\n)$-Calogero
 algebra is the $SU(2)$-polynomial (cubic) algebra\cite{2body}, i.e., 
$[X_{2,0},X_{0,2}]=P_3({\cal N}_1,{\cal \br N})$. In this case, the  algebra 
${\cal C}_2(\n)$ can be linearized to the ordinary $SU(2)$ algebra 
owing to the fact that there are two independent, uncoupled oscillators $B_1$ 
and $A_2$, which can be mapped to two ordinary Bose oscillators\cite{mmp}.
The $SU(2)$ generators are
\baa\label{su2}
J_+&=&\frac{1}{4\sqrt{({\cal N}_1-1)({\cal \br N}+1+2\n)}}B_1^{\dagger 2}A_2,\nn
\left(J_+\right)^{\dagger}&=&
J_-=A_2^{\dagger}B_1^2\frac{1}{4\sqrt{({\cal N}_1-1)({\cal \br N}+1+2\n)}},\nn
J_0&=&\frac{1}{16}\left(\frac{1}{({\cal N}_1-1)}B_1^{\dagger 2}B_1^2-
\frac{4}{({\cal \br N}-1+2\n)} A_2^{\dagger}A_2\right)=\frac{1}{4}
\left({\cal N}_1-{\cal \br N}\right),
\ea
satisfying $[J_+,J_-]=2J_0,\;[J_0,J_{\pm}]=\pm J_{\pm}$.
The generators $J_+$ and $J_-$ are hermitian conjugates to each other and in 
this respect differ from the construction done in Ref.\cite{ind}.
For $N=3$, the ${\cal C}_3(\n)$ algebra in Eqs.(\ref{n},\ref{o}) 
is the same as in Ref.\cite{3body}. One can easily find the 
exact correspondence 
using Eq.(\ref{corr}). 
For general $N$, our construction can be viewed as a generalization of the
polynomial algebras for $N=2$\cite{2body} and $N=3$\cite{3body,Sn},
using the algebra ${\cal B}_N(\n)$.

Finally,  it would be interesting to construct some different sets of 
generators of dynamical algebra, and  to discuss the relation 
between   bosonsic realization of the $SU(N)$ algebra and the   
${\cal C}_N(\n)$ algebra, generalizing  Eq.(\ref{su2}) for any $N$. 
We plan to answer some of these open questions in the forthcoming 
publication\cite{inprep}. 

Acknowledgment

We would like to thank M. Milekovi\'c and D. Svrtan for useful
discussions.
this work was supported by the Ministry of Science and Technology of the
 Republic of Croatia under contract No. 00980103.

\end{document}